\documentclass[12pt]{article}
\usepackage{amsfonts}

\newcommand{\sect}[1]{\setcounter{equation}{0}\section{#1}}
\newcommand{\subsect}[1]{\subsection{#1}}

\def\be{\begin{equation}}
\def\ee{\end{equation}}
\def\bea{\begin{eqnarray}}
\def\eea{\end{eqnarray}}

\def\1{\'{\i}}                           
\def\R{{\mathbb R}}

\def\jp{J_+}
\def\jm{J_-}
\def\jj{J_3}

\def\para{\omega^2}

\def\kk{K}

\def\s{{\sigma}} 

\def\otra{b}

\def\co{\Delta}

\def\w{\omega}

\def\j{{\bar\sigma}} 
\def\s{{\sigma}} 

\def\>#1{{\bf #1}}

\def\tfrac#1#2{ {\scriptstyle { \frac {#1}{#2}}}}         
\def\pois#1#2{\left\{ {#1},{#2} \right\}}

\begin{document}

\thispagestyle{empty}
\hfill\ 
\vspace{2.5cm}

\begin{center} {\LARGE{\bf{Integrable deformations of oscillator
}}} 

{\LARGE{\bf{chains from quantum algebras}}} 

\vspace{1.4cm} 

{\sc \'Angel Ballesteros}\footnote{e-mail: angelb@ubu.es. Phone:
(34) 947 258894. Fax: (34) 947 258831.} {\sc and
Francisco J. Herranz}\vspace{.2cm}\\ {\it Departamento de F\1sica,
Universidad de Burgos\\ Pza. Misael Ba\~nuelos s.n., 09001-Burgos,
Spain} 
\vspace{.3cm}

\end{center} 
  
\bigskip

\begin{abstract} 
A family of completely integrable nonlinear  deformations of
systems of $N$ harmonic oscillators are constructed from the
non-standard quantum deformation of the $sl(2,\R)$ algebra.
Explicit expressions for all the associated integrals of motion
are given, and the long-range nature of the interactions
introduced by the deformation is shown to be linked to the
underlying coalgebra structure. Separability and
superintegrability properties of such systems are analysed, and
their connection with classical angular momentum chains is used
to construct a non-standard integrable deformation of the XXX
hyperbolic Gaudin system.
\end{abstract} 

\vspace{1.4cm}

\vfill\eject

%%%%%%%%%%%%%%%%%%%%%%%%%%%%%%%%%%%%%%%%%%%%%%%

\sect{Introduction}

The construction of integrable systems is an outstanding
application of Lie algebras in both  classical and quantum
mechanics \cite{OP,Per}. In fact, the very definition of
integrability is based on the concept of involutivity of the
conserved quantities with respect to a (either Poisson or
commutator) Lie bracket. During last years, many new results
concerning `quantum' deformations of Lie algebras and groups have
been obtained (see, for instance \cite{CP}), and this work has
extended in many different directions the original deformations
that appeared in the context of (Classical and Quantum) Inverse
Scattering methods \cite{Fad}. Therefore, the question concerning
whether all these new nonlinear algebraic structures can be
connected in a systematic way with the integrability properties
of a certain class of dynamical systems arises as a keystone for
future developments in the subject.

The aim of this paper is to answer this question in the
affirmative  by explicitly constructing some $N$-dimensional
systems through the general and systematic construction of
integrable systems from coalgebras that has been introduced in
\cite{BR}. Such procedure is essentially based on the role that
the coalgebra structure, i.e., the existence of an homomorphism
$\Delta:A\rightarrow A\otimes A$ defined on a one-particle
dynamical algebra $A$, plays in the propagation of the
integrability from  the one-body problem to a general $N$-particle
Hamiltonian with  coalgebra symmetry. In this framework, quantum
algebras (which  are just coalgebra deformations) can be
interpreted as dynamical symmetries that generate in a direct way
a large class of integrable deformations. In order to extract the
essential properties of the systems associated to quantum
algebras, we shall concentrate here on the explicit construction
and analysis of integrable deformations of (classical mechanical)
oscillator chains obtained from quantum
$sl(2,\R)$  coalgebras. We recall
that  quantum algebra deformations of $sl(2,\R)$ are
basic in quantum  algebra theory and can be found, for instance,
in
\cite{CP}.

In the next Section we briefly   summarize the general
construction of \cite{BR} and fix the notation. Section 3 deals
with oscillator chains obtained from the non-standard  quantum  
$sl(2,\R)$ coalgebra \cite{Demidov}--\cite{Ohn} through a linear 
Hamiltonian of the type ${\cal H}=\jp + \alpha\,\jm$. This 
deformation
  can be interpreted either as a  direct algebraic implementation
of a certain type of long-range interaction or, equivalently, in
relation with a certain integrable perturbation of the motion of
a  particle under any central potential in the $N$-dimensional
Euclidean space. Through these examples we will show an intrinsic
connection between quantum deformations  and nonlinear
interactions depending on the momenta. Next, the construction
of  anharmonic chains is  studied, thus showing the  number of
integrable systems that can be easily derived by following the
present approach with different choices for the generating
Hamiltonian
${\cal H}$.

The problem of separation of variables of  these oscillator chains
is analysed in Section 4. As a result, it is shown that the
integrable derformation introduced in Section 3 is not separable.
However some other choices for ${\cal H}$ lead to St\"ackel
systems, thus preserving separability and superintegrability after
deformation.  In Section 5 we also present the direct
relationship between these
$sl(2,\R)$ oscillator chains and classical spin models, and we
explicitly construct the non-standard deformation of the
classical XXX hyperbolic Gaudin system
\cite{Gau}--\cite{Cal}. As it happened with the standard
deformation, the non-standard one generates a complicated
variable range \cite{Inoz} integrable interaction. Some final
remarks and comments close the paper.

%%%%%%%%%%%%%%%%%%%%%%%%%%%%%%%%%%%%%%%%%%%%%%%

\section{From coalgebras to integrable Hamiltonians}

The main result of \cite{BR} can be summarized as follows: Any
coalgebra $(A,\co)$  with Casimir  element $C$ can be considered
as the generating symmetry that, after choosing a non-trivial
representation, gives rise to a large family
of integrable systems in a  systematic way. Here, we shall
consider classical mechanical systems and, consequently, we shall
make use of Poisson  realizations $D$ of Lie and quantum algebras
of the form $D:A\rightarrow C^\infty (q,p)$. However, we recall
that the formalism is also directly applicable to quantum
mechanical systems.

Let $(A,\Delta)$ be a (Poisson) coalgebra with generators $X_i$
 $(i=1,\dots,l)$ and Casimir element $C(X_1,\dots,X_l)$.
Therefore, the coproduct $\Delta:A\rightarrow A\otimes A$ is a
Poisson map with respect to the usual   Poisson bracket on
$A\otimes A$:
\be
\pois{X_i\otimes X_j}{X_r\otimes X_s}_{A\otimes  A}=
\{X_i, X_r\}\otimes X_j  X_s +
 X_i  X_r \otimes \{X_j, X_s \}  .
\ee

Let us consider the $N$-th coproduct $\co^{(N)}(X_i)$ of the
generators 
\be
\co^{(N)}:A\rightarrow A\otimes A\otimes \dots^{N)}\otimes A 
\ee
which is obtained (see \cite{BR}) by applying recursively the
two-coproduct $\co^{(2)}\equiv \co$ in the form
\be
\co^{(N)}:=(id\otimes id\otimes\dots^{N-2)}\otimes id\otimes
\co^{(2)})\circ\co^{(N-1)}.
\label{fl}
\ee
By taking into account that  the $m$-th coproduct $(m\leq N)$ of
the Casimir $\co^{(m)}(C)$ can be embedded into the tensor
product of
$N$ copies of $A$ as
\be
\co^{(m)}: A\rightarrow \{A\otimes A\otimes \dots^{m)}\otimes A\}
\otimes \{1 \otimes 1\otimes \dots^{N-m)}\otimes 1\} 
\ee
it can be shown that  
\be
\pois{\co^{(m)}(C)}{\co^{(N)}(X_i)}_{A\otimes
A\otimes\dots^{N)}\otimes A}=0 \qquad
i=1,\dots,l \quad m=2,\dots,N .
\label{za}
\ee

With this in mind it can be proven \cite{BR} that, if ${\cal
H}$ is an {\it arbitrary}  (smooth) function of the generators of
$A$, the $N$-particle Hamiltonian defined on
$A\otimes A\otimes\dots^{N)}\otimes A$ as the $N$-th coproduct of
${\cal H}$
\be
H^{(N)}:=\co^{(N)}({\cal{H}}(X_1,\dots,X_l))=
{\cal{H}}(\co^{(N)}(X_1),\dots,\co^{(N)}(X_l)),
\label{htotg}
\ee
fulfils
\be
\pois{C^{(m)}}{H^{(N)}}_{A\otimes
A\otimes\dots^{N)}\otimes A}=0 \qquad m=2,\dots,N 
\label{za1}
\ee
where the $N-1$ functions $C^{(m)}$ ($m=2,\dots,N$) are defined
through the coproducts of the Casimir $C$
\be
C^{(m)}:= \co^{(m)}(C(X_1,\dots,X_l))=
C(\co^{(m)}(X_1),\dots,\co^{(m)}(X_l))
\label{Ctotg}
\ee
and all the integrals of motion $C^{(m)}$ are in involution
\be
\pois{C^{(m)}}{C^{(n)}}_{A\otimes
A\otimes\dots^{N)}\otimes A}=0 \qquad  m,n=2,\dots,N.
\label{cor3}
\ee

Therefore, once a realization of $A$ on a one-particle phase
space is given, the $N$-particle Hamiltonian $H^{(N)}$ will be a
function of $N$ canonical pairs  $(q_i,p_i)$ and is, by
construction, completely  integrable with respect to the usual
Poisson bracket 
\be
\{f,g\}=\sum_{i=1}^N\left(\frac{\partial f}{\partial q_i}
\frac{\partial g}{\partial p_i}
-\frac{\partial g}{\partial q_i} 
\frac{\partial f}{\partial p_i}\right) .
\label{poisbra}
\ee
Furthermore, its  integrals of motion will be given by the
$C^{(m)}$ functions, all of them functionally independent since
each of them depends on the first $m$ pairs $(q_i,p_i)$ of
canonical coordinates. 

In particular, this result   can be applied to universal
enveloping algebras of Lie algebras $U(g)$ \cite{BCR}, since they
are always endowed with a natural (primitive)  Hopf algebra
structure of the form 
\be
\co(X_i)= X_i\otimes 1+1\otimes X_i 
\label{primi}
\ee
 being $X_i$  any generator  of $g$. Moreover, since quantum
algebras are also (deformed) coalgebras $(A_z,\Delta_z)$, any
function of the generators of  a given quantum algebra with
Casimir element $C_z$ will provide, under a chosen deformed
representation, a completely integrable Hamiltonian.

%%%%%%%%%%%%%%%%%%%%%%%%%%%%%%%%%%%%%%%%%%%%%%%%%%%%%%%

\sect{Oscillator chains from $sl(2,\R)$ coalgebras}

The obtention of integrable  oscillator chains by using the
previous approach can be achieved by selecting Poisson 
coalgebras $(A,\co)$ such that the one-dimensional harmonic
oscillator Hamiltonian with angular frecuency $\w$ (and unit
mass) can be written as the phase space representation $D$ of a
certain function ${\cal H}$ of the generators of $A$:  
\be
H=D ({\cal H})=p^2 + \w^2 q^2.
\label{oh}
\ee
It is well-known that   $sl(2,\R)$ can be considered as a
dynamical algebra for $H$. Hence, when deformations of $sl(2,\R)$
coalgebras are considered, a big class of new
integrable deformations of oscillator chains can be obtained.

%%%%%%%%%%%%%%%%%%%%%%%%%%%%%%%%%%%%%%%%%%%%%%%%%%%%%%%

In particular, let us introduce  the Poisson bracket analogue of
the non-standard deformation of
$sl(2,\R)$
\cite{Ohn} given by
\be 
\{\jj,\jp\}=2 \jp \cosh z\jm  \qquad 
 \{\jj,\jm\}=-2\frac {\sinh z\jm}{z}\qquad
\{\jm,\jp\}=4 \jj   
\label{gb}
\ee 
where $z$ is the deformation parameter. A Poisson coalgebra
structure, $(U_z(sl(2,\R)),\Delta_z)$,  is obtained by means of
the following coproduct
\bea 
&&\Delta_z(\jm)=  \jm \otimes 1+
1\otimes \jm \cr
 &&\Delta_z(\jp)=\jp \otimes e^{z \jm} + e^{-z \jm} \otimes
\jp \cr
 &&\Delta_z(\jj)=\jj \otimes e^{z \jm} + e^{-z \jm} \otimes \jj .
\label{ga}
\eea
The corresponding Casimir function reads
\be
{\cal C}_z=\jj^2  - \frac {\sinh z\jm}{z} \jp .
\label{gc}
\ee
A one-particle  phase space realization  $D_z$ of (\ref{gb}) is
given by:
\bea 
&&{\tilde f}_-^{(1)}=D_z(\jm)=q_1^2\cr
&&{\tilde f}_+^{(1)}= D_z(\jp)=\frac {\sinh z q_1^2}{z q_1^2}
p_1^2 +
\frac{z \otra_1}{\sinh z q_1^2} \label{gd}\\
&&{\tilde f}_3^{(1)}=D_z(\jj)=\frac {\sinh z q_1^2}{z q_1^2} q_1
p_1  
\nonumber
\eea
where $\otra_1$  is a  real constant that labels the
representation through the  Casimir: $C_z^{(1)}=D_z({\cal
C}_z)=-\otra_1$.

Let us now consider  the dynamical generator
\be
{\cal H}= \jp+\para  \jm .
\label{hsl}
\ee
 Under (\ref{gd}), we obtain a new 
deformation of the oscillator Hamiltonian  (\ref{oh}) including a
deformed centrifugal term governed by the parameter $\otra_1$:
\be
H^{(1)}_z=D_z({\cal H})={\tilde f}_+^{(1)}+\para  {\tilde
f}_-^{(1)}= \frac {\sinh z q_1^2}{z q_1^2} 
p_1^2  +\para q_1^2 +\frac{z \otra_1}{\sinh z q_1^2} . 
\label{ge}
\ee

Now, we follow the constructive   method of section 2 and derive
the two-particle phase space realization from the   coproduct
(\ref{ga}) and two copies of the realization (\ref{gd}):
\bea
 &&{\tilde f}_-^{(2)}=(D_z\otimes
D_z)(\Delta_z(\jm))=q_1^2+q_2^2\cr &&{\tilde
f}_+^{(2)}=(D_z\otimes D_z)(\Delta_z(\jp))\cr &&\qquad\qquad=
 \left( \frac {\sinh z q_1^2}{z q_1^2}  p_1^2 +\frac{z
\otra_1}{\sinh z q_1^2} \right) e^{z q_2^2} +
 \left(\frac {\sinh z q_2^2}{z q_2^2}  p_2^2  +\frac{z
\otra_2}{\sinh z q_2^2} \right) e^{-z q_1^2}\cr
&&{\tilde f}_3^{(2)}=(D_z\otimes D_z)(\Delta_z(\jj))=
\frac {\sinh z q_1^2}{z q_1^2 }  q_1 p_1   e^{z q_2^2} +
\frac {\sinh z q_2^2}{z q_2^2 }  q_2 p_2  e^{-z q_1^2} .
\label{gf}
\eea
It can be easily checked that these functions  close again  the
deformed algebra (\ref{gb})  under the usual Poisson bracket
$\pois{q_i}{p_j}=\delta_{i\,j}$.

By following (\ref{htotg}), the two-particle Hamiltonian 
will be given by the realization of the coproduct of ${\cal H}$:
   $H^{(2)}_z=(D_z\otimes D_z)(\Delta_z({\cal H}))= {\tilde
f}_+^{(2)}+\para {\tilde f}_-^{(2)}$; it reads 
\bea 
&& H^{(2)}_z = \frac {\sinh z
q_1^2}{z q_1^2}  p_1^2  e^{z q_2^2} +
\frac {\sinh z q_2^2}{z q_2^2}  p_2^2  e^{-z q_1^2} +\para
  ( q_1^2+q_2^2  ) \cr
&&\qquad\qquad  +\frac{z \otra_1}{\sinh z
q_1^2}   e^{z q_2^2} + \frac{z \otra_2}{\sinh z
q_2^2}   e^{-z q_1^2} .
\label{gg}
\eea 
The  coproduct for the Casimir, $C^{(2)}_z= (D_z\otimes
D_z)(\Delta_z({\cal C}_z))$,  leads to the following integral of
the motion for (\ref{gg}):
\bea
&& C^{(2)}_z =-\frac {\sinh z q_1^2
}{z q_1^2 } \,
\frac {\sinh z q_2^2}{z q_2^2} 
\left({q_1}{p_2} - {q_2}{p_1}\right)^2
 e^{-z q_1^2}e^{z q_2^2} -(\otra_1 e^{2z q_2^2} +\otra_2 e^{-2z
q_1^2})\cr &&\qquad\qquad -\left( \otra_1 \frac {\sinh z
q_2^2}{\sinh z q_1^2} + \otra_2 \frac {\sinh z q_1^2}{\sinh z
q_2^2} \right) e^{-z q_1^2}e^{z q_2^2}.
\eea

 The $N$-dimensional generalization for this system follows from
the realization of the coalgebra on an $N$-dimensional phase
space. In general, an $m$-dimensional phase space realization is
obtained through the tensor product of $m$ copies of (\ref{gd})
applied onto the $m$-th deformed coproduct (\ref{fl}), which is
in turn induced from the two-body coproduct (\ref{ga}) (see
 \cite{BR} for an explicit example). In our case, this
construction gives
\bea  
&& 
{\tilde f}_-^{(m)}= \sum_{i=1}^m q_i^2\cr 
&& {\tilde f}_+^{(m)}=\sum_{i=1}^m
 \left( \frac {\sinh z q_i^2}{z q_i^2}  p_i^2  +\frac{z
\otra_i}{\sinh z q_i^2} \right)  e^{z \kk_i^{(m)}(q^2) }\cr
&&{\tilde f}_3^{(m)}=\sum_{i=1}^m
\frac {\sinh z q_i^2}{z q_i^2}  q_ip_i  e^{z \kk_i^{(m)}(q^2) }  
\label{gi}
\eea
where the $K$-functions that we will use hereafter are defined by
\bea
 \kk_i^{(m)}(x)& =&  - \sum_{k=1}^{i-1}  x_k+ 
\sum_{l=i+1}^m   x_l
\label{zzd}\\
 \kk_{ij}^{(m)}(x) & =&\kk_i^{(m)}(x)+ \kk_j^{(m)}(x)\cr
 & =&   - 2\sum_{k=1}^{i-1}   x_k -   x_i +   x_j +
2\sum_{l=j+1}^m   x_l  \qquad i<j .
\label{zze} 
\eea
From now on, any sum 
 defined on an empty set of indices will be assumed to be zero.
For instance, $\kk_1^{(3)}(x)=x_2+x_3$, $\kk_2^{(3)}(x)=-x_1+x_3$
and
$\kk_3^{(3)}(x)=-x_1-x_2$.

As a consequence, the $N$-dimensional Hamiltonian associated to
the dynamical generator (\ref{hsl}) is just 
\be
H^{(N)}_z =  {\tilde f}_+^{(N)}+\para {\tilde f}_-^{(N)}=
\sum_{i=1}^N
 \left( \frac {\sinh z q_i^2}{z q_i^2}  p_i^2  +\frac{z
\otra_i}{\sinh z q_i^2} \right)  e^{z \kk_i^{(N)}(q^2) } +
\para\,\sum_{i=1}^N q_i^2.
\label{ddr}
\ee
This characterizes a chain of interacting oscillators where 
the long-range nature of the coupling  introduced by the
 deformation is encoded through the functions $\kk_i^{(N)}(q^2)$.
The following $N-1$ integrals of motion are deduced from the
$m$-th coproducts of the Casimir $(m=2,\dots,N)$: 
\bea 
&&  C^{(m)}_z=- \sum_{i<j}^m 
\frac {\sinh z q_i^2}{z  q_i^2}
\frac {\sinh z q_j^2}{z q_j^2}
\left({q_i}{p_j} - {q_j}{p_i}\right)^2  e^{ z
 \kk_{ij}^{(m)}(q^2)} - \sum_{i=1}^m \otra_i  e^{2 z
\kk_{i}^{(m)}(q^2)}\cr &&\qquad\qquad - \sum_{i<j}^m 
\left( \otra_i \frac {\sinh z q_j^2}{\sinh z q_i^2}
+ \otra_j \frac {\sinh z q_i^2}{\sinh z q_j^2} \right)  e^{ z
\kk_{ij}^{(m)}(q^2)} .
\label{gn}
\eea 
 We point out that the following property is useful in the
previous computations:
 \be \frac {\sinh(z\sum_{i=1}^m   x_i)}z = \sum_{i=1}^m  
\frac {\sinh z   x_i}z  e^{ z \kk_{i}^{(m)}(x)} .
\label{zzf}
\ee

The underformed counterpart of the above systems can be directly
obtained by applying the limit $z\to 0$ in all the expressions
that we have just deduced. In particular, the  Poisson coalgebra
$(U(sl(2,\R)),\co)$ is defined by the   Lie--Poisson algebra
\be
\{\jj,\jp\}=2\jp\qquad 
\{\jj,\jm\}=-2\jm\qquad \{\jm,\jp\}=4\jj 
\label{fb}
\ee
 together with the primitive coproduct (\ref{primi}), and the
Casimir
 is  ${\cal C}=\jj^2 - \jm \jp$. Once the limit $z\to 0$ is
computed, the  deformed phase space realization (\ref{gi}),
 Hamiltonian (\ref{ddr}) and  integrals of motion (\ref{gn})
reduce to
\be 
 f_-^{(m)}= 
\sum_{i=1}^m   q_i^2\qquad
 f_+^{(m)} =\sum_{i=1}^m  
\biggl(p_i^2 + \frac{\otra_i}{q_i^2}\biggr) \qquad
 f_3^{(m)}=
\sum_{i=1}^m q_i p_i 
\label{fi}
\ee 
\be 
H^{(N)}=\sum_{i=1}^N \biggl( p_i^2 +\para q_i^2+
\frac{\otra_i}{q_i^2}\biggr)  
\label{fj}
\ee
\be  
  C^{(m)}=  -\sum_{i<j}^m ({q_i}{p_j} - {q_j}{p_i})^2 -
\sum_{i<j}^m\left(
\otra_i\frac{q_j^2}{q_i^2}+\otra_j\frac{q_i^2}{q_j^2}\right)
-\sum_{i=1}^m \otra_i .
\label{fk}
\ee 
Consequently, the non-deformed Poisson coalgebra
 $(U(sl(2,\R)),\co)$  provides an uncoupled chain of $N$ harmonic
oscillators (\ref{fj}) (all of them with the same frecuency) with
centrifugal terms. We remark  that the (well-known \cite{OP})
complete integrability of
$ H^{(N)}$ is obtained  directly from its underlying
coalgebra symmetry. Moreover, if the centrifugal terms disappear
($\otra_i=0$), the integrals $C^{(m)}$  are just the quadratic
 Casimirs of the $so(m)$ algebras with $m=2,\dots,N$. It is also 
a classical result that the Hamiltonian (\ref{fj}) is $so(N)$
invariant, since it can be interpreted as the one for a particle
moving on the $N$-dimensional Euclidean space under the central
potential $\para\,r^2$. We stress that all these known
considerations are deduced in a straightforward way from the
coalgebra symmetry of the model.  

%%%%%%%%%%%%%%%%%%%%%%%%%%%%%%%%%%%%%%%%%%%%%%%%%%%%%%%

\subsect{A class of integrable anharmonic chains}

 It is also possible to consider the non-deformed Poisson
coalgebra
 $(U(sl(2,\R)),\co)$ and a more general dynamical Hamiltonian 
than (\ref{hsl}) of the form
\be
{\cal H}=\jp+ {\cal F}(\jm)
\label{anh}
\ee
where ${\cal F}(\jm)$ is an arbitrary smooth  function of $\jm$.
 The formalism ensures that the corresponding system constructed
from (\ref{anh}) is also completely integrable, since ${\cal H}$
could be {\it any function} of the coalgebra generators.
Explicitly, this means that any $N$-particle Hamiltonian of the
form
\be 
 H^{(N)}
=    f_+^{(N)} +{\cal F}(f_-^{(N)}) 
=\sum_{i=1}^N \biggl( p_i^2 + \frac{\otra_i}{q_i^2}  \biggr)+
{\cal F}\biggl(\sum_{i=1}^N q_i^2\biggr)
\label{anha}
\ee
 is completely integrable, and (\ref{fk}) are its integrals of
motion. Obviously, in the case $\otra_i=0$, this is a well-known
result, since (\ref{anha}) is just the Hamiltonian describing the
motion of a particle in an $N$-dimensional Euclidean space under
the action of a central potential. The linear function ${\cal
F}(\jm)=\para\,\jm$ leads to the previous harmonic case, and the
quadratic one ${\cal F}(\jm)=\jm^2$ would give us an interacting
chain of quartic oscillators. Further definitions of the function
${\cal F}$ would give rise to many other anharmonic chains, all
of them sharing the same dynamical symmetry and the same
integrals of the motion.

 Moreover, the corresponding integrable deformation of
 (\ref{anha}) is provided by a realization of (\ref{anh}) in terms
of (\ref{gi}):
\be 
 H_z^{(N)} 
=\sum_{i=1}^m
 \left( \frac {\sinh z q_i^2}{z q_i^2}  p_i^2  +\frac{z
\otra_i}{\sinh z q_i^2} \right)  e^{z \kk_i^{(m)}(q^2) }
 + {\cal F}\left(\sum_{i=1}^N q_i^2\right)
\label{anhaz}
\ee
 and (\ref{gn}) are again the associated integrals. This example
shows clearly the number of different systems that can be obtained
through the same coalgebra, and the need for a careful
 inspection of known integrable systems in order to investigate
their possible coalgebra symmetries.

%%%%%%%%%%%%%%%%%%%%%%%%%%%%%%%%%%%%%%%%%%%%%%%%%%%%%%%

\sect{Separation of variables and superintegrability}

It is clear that  $H^{(N)}$ (\ref{fj}) is the Hamiltonian of a
Liouville system  \cite{Per}, so that we find  another set of
integrals of  motion in involution given by
\be
I_i=p_i^2 +\para q_i^2+
\frac{\otra_i}{q_i^2}- \frac {H^{(N)}}{N}\qquad
i=1,\dots,N .
\label{sa}
\ee
Amongst these quantities, only $N-1$ are functionally independent
($\sum_{i=1}^N I_i=0$) and, obviously, the following
Hamilton--Jacobi equation admits a separable solution:
\be
H^{(N)}(q_1,\dots,q_N;p_1,\dots,p_N)=E\qquad p_i=\frac{\partial
W}{\partial q_i}\qquad W\equiv W(q_1,\dots,q_N)=\sum_{i=1}^N
W_i(q_i) .
\label{sb}
\ee
The integrals of motion $I_i$ are independent with respect to the 
$C^{(m)}$ (\ref{fk}) and, in general, $\{C^{(m)},I_i\}\ne 0$. 
Hence, $H^{(N)}$ is a superintegrable system.

Unlike $H^{(N)}$,  the deformed Hamiltonian $H^{(N)}_z$
 (\ref{ddr}) does no longer define a  Liouville system.  In order
to  analyse if $H^{(N)}_z$ admits separation of variables 
we recall the general criterion for the separability problem of 
 the  Hamilton--Jacobi equation (\ref{sb}): This equation is
 separable if the $N$-particle Hamiltonian $H$ verifies the
following set of $N(N-1)/2$ equations \cite{Per}:  
\be
 \frac{\partial H}{\partial p_i}\,\frac{\partial H}{\partial
p_j}\,
\frac{\partial^2 H}{\partial q_i\partial q_j}-
 \frac{\partial H}{\partial p_i}\,\frac{\partial H}{\partial
q_j}\,
\frac{\partial^2 H}{\partial q_i\partial p_j}-
 \frac{\partial H}{\partial q_i}\,\frac{\partial H}{\partial
p_j}\,
\frac{\partial^2 H}{\partial p_i\partial q_j}+
 \frac{\partial H}{\partial q_i}\,\frac{\partial H}{\partial
q_j}\,
\frac{\partial^2 H}{\partial p_i\partial p_j}=0
\label{sc}
\ee
where $i,j=1,\dots,N$ and $i<j$.

If we  consider the two-particle Hamiltonian
 $H^{(2)}_z$ (\ref{gg}), it can be checked that the single
equation
  (\ref{sc}) is not satisfied. This is due to the long-range
nature of the deformation  characterized by the $K$-functions:
If  we make
$K_i^{(2)}=0$  then the  general criterion of separability is
fulfilled (the same happens in higher dimensions).

However, the coalgebra construction allows for
an infinite family of  completely integrable deformations, all of
 them sharing the same integrals of motion. With this in mind, it
 is natural  to study whether some `modifications' of the  initial
 dynamical generator $\cal H$ (\ref{hsl}) (whose non-deformed
limit ($z\to 0$) lead again to $H^{(N)}$ (\ref{fj}))
enable us to find a separable deformed Hamiltonian. 
As a first step we analyse the 
 two-particle case (\ref{gf}) and consider two new `candidates'
for 
$\cal H$:
\be 
{\cal H}_1= \jp e^{\alpha_1 z \jm} +\para  \jm e^{\beta_1 z \jm}
\  \Rightarrow \ 
H_{1,z}^{(2)}={\tilde f}_+^{(2)}
 e^{\alpha_1 z {\tilde f}_-^{(2)}} +\para {\tilde f}_-^{(2)}
 e^{\beta_1 z {\tilde f}_-^{(2)}}
\label{sd}
\ee
\be
 {\cal H}_2= \jp e^{\alpha_2 z \jm} +\para 
\left(\frac{e^{\beta_2 z
\jm}-1}{\beta_2 z}\right)
\   \Rightarrow \ 
H_{2,z}^{(2)}={\tilde f}_+^{(2)}
 e^{\alpha_2 z {\tilde f}_-^{(2)}} +\para  \left(\frac{
 e^{\beta_2 z {\tilde f}_-^{(2)}} -1 }{\beta_2 z}\right)
\label{se}
\ee
where $\alpha_1$,  $\alpha_2$, $\beta_1$ and $\beta_2$ are real
constants; note that under the limit $z\to 0$ we recover, in both
cases, the non-deformed Hamiltonian $H^{(2)}$.
 If we impose the equation (\ref{sc}) to be fulfilled, then we
find two solutions for each new Hamiltonian:
\bea 
 H_{1,z}^{(2)}:  &&  (i)\quad \alpha_1=1\quad \w=0\qquad\  
(ii)\quad
\alpha_1=-1\quad \w=0
\label{sf}\\
H_{2,z}^{(2)}:  &&  (i)\quad \alpha_2=1\quad \beta_2=2\qquad 
(ii)\quad
\alpha_2=-1\quad \beta_2=-2
\label{sg}
\eea
Since the two  solutions associated to   $H_{1,z}^{(2)}$ arise as
 particular cases of those corresponding to $H_{2,z}^{(2)}$ once
the frecuency $\w$ vanishes,  we only consider the latter. We
stress that the two solutions (\ref{sg}) do not only provide
separable Hamiltonian systems in the Hamilton--Jacobi equation
(\ref{sb}), but  they are  also
 St\"ackel systems \cite{Per}. Furthermore, this property can be
generalised to the arbitrary $N$-particle case. 
In the following we construct the St\"ackel description for the
first solution of (\ref{sg}).

The dynamical generator we start with is given by
\be
{\cal H}= \jp e^{ z \jm} +\para  \left(\frac{e^{2 z
\jm}-1}{2 z}\right) .
\label{sh}
\ee
 By introducing the realization (\ref{gi}) we obtain the
Hamiltonian
\bea
&&H^{(N)}_z =  \sum_{i=1}^N \frac {\sinh z q_i^2}{z q_i^2}\, 
e^{z q_i^2}\exp\left\{ {2z \sum_{k=i+1}^N q_k^2} \right\}
\left( p_i^2  +  \otra_i \left(\frac{z q_i}{\sinh z
q_i^2}\right)^2  \right)
\cr
&&\qquad\qquad\qquad\qquad
+\para  \left(\frac{\exp\left\{2 z\sum_{j=1}^N q_j^2\right\}-1}{2
z}\right)
\label{si}
\eea
which has the form of a St\"ackel system
\be
H^{(N)}_z=\sum_{i=1}^N a_i(q_1,\dots,q_N) \left(\frac 12 p_i^2 +
U_i(q_i)\right)
\label{sj}
\ee
provided that 
\bea
&&a_i(q_1,\dots,q_N)=
2\,\frac {\sinh z q_i^2}{z q_i^2}\,
e^{z q_i^2}\exp\left\{ {2z \sum_{k=i+1}^N q_k^2}\right\}\qquad
i=1,\dots,N\cr
&&U_1(q_1)=
\frac {\otra_1}{2} \left(\frac{z q_1}{\sinh z
q_1^2}\right)^2 +\frac{\w^2}{4z}\, e^{z q_1^2}
\,\frac {z q_1^2}{\sinh z q_1^2}\cr
&&U_i(q_i)=\frac {\otra_i}{2} \left(\frac{z q_i}{\sinh
z q_i^2}\right)^2 \qquad i=2,\dots,N-1\cr
&&U_N(q_N)=
\frac {\otra_N}{2} \left(\frac{z q_N}{\sinh z
q_N^2}\right)^2 -\frac{\w^2}{4z}\, e^{-z q_N^2}
\,\frac {z q_N^2}{\sinh z q_N^2} .
\label{sk}
\eea
We recall that St\"ackel's theorem \cite{Per} claims that a
 system with a Hamiltonian of the form (\ref{sj}) admits
separation of variables in the   Hamilton--Jacobi equation
(\ref{sb}) if and only if there exists an $N\times N$ matrix $B$
whose entries
$b_{ij}$ depend only on $q_j$, and such that
\be
\mbox{det}\,B\ne 0\qquad \sum_{j=1}^N b_{ij}(q_j)
a_j(q_1,\dots,q_N)=\delta_{i1} .
\label{sl}
\ee
These requirements are satisfied by our Hamiltonian (\ref{si});
 the non-zero elements of $B$ and its determinant are
found to be
\bea 
&&b_{1N}(q_N)=\frac {z q_N^2}{2\sinh z q_N^2}\,e^{-z q_N^2}
\qquad
b_{i\,i-1}(q_{i-1})=\frac {z q_{i-1}^2}{\sinh z q_{i-1}^2}\,e^{-z
q_{i-1}^2}\cr
&&
b_{ii}(q_{i})=-\frac {z q_{i}^2}{\sinh z q_{i}^2}\,e^{z q_{i}^2}
\qquad
i=2,\dots,N 
\label{sm}
\eea 
\be
\mbox{det}\,B =\frac 12 \prod_{i=1}^N
\frac {z q_{i}^2}{\sinh z q_{i}^2}\,e^{-z q_{i}^2}.
\label{sn}
\ee
The St\"ackel's theorem gives $N$ functionally independent
integrals of motion  in involution which have the form
\be
I_j=\sum_{i=1}^N a_{ij} \left(\frac 12 p_i^2 +
U_i(q_i)\right)\qquad j=1,\dots,N
\label{so}
\ee
 where $a_{ij}$ are the elements of the inverse matrix to $B$.
Then
$a_{i1}=a_i$, so that the first integral $I_1$ is just the
Hamiltonian. In our case, the non-zero functions
$a_{ij}$ read
\bea
&&a_{i1}=2\,\frac {\sinh z q_{i}^2}{z q_{i}^2}\,e^{z q_{i}^2}
 \exp\left\{ {2z \sum_{k=i+1}^N q_k^2}\right\}\qquad
i=1,\dots,N\cr &&a_{ij}= \frac {\sinh z q_{i}^2}{z q_{i}^2}\,e^{z
q_{i}^2}
 \exp\left\{ {2z \sum_{k=i+1}^{j-1} q_k^2}\right\}\qquad
i=1,\dots,N
\quad i<j .
\label{sp}
\eea
Consequently, we have proven that besides the $N-1$ integrals of
motion  $C^{(m)}_z$ (\ref{gn}), the Hamiltonian (\ref{si}) has
another set of $N-1$ conserved quantities given by (\ref{so}):
\be
I_j=\sum_{i=1}^{j-1} 
\frac {\sinh z q_{i}^2}{2z q_{i}^2}\,e^{z q_{i}^2}
\exp\left\{ {2z \sum_{k=i+1}^{j-1} q_k^2}\right\}
\left(p_i^2+  {\otra_i} \left(\frac{z q_i}{\sinh
z q_i^2}\right)^2\right)+
\frac{\w^2}{4z} \exp\left\{ {2z \sum_{k=1}^{j-1} q_k^2}\right\}
\label{sq}
\ee
with $j=2,\dots,N$. Note that the non-deformed limit for $I_j$ has
to be computed as $\lim_{z\to 0}(I_j -\frac {\w^2}{4z})$ in order
to avoid divergencies.
 
The integrals of motion $I_j$ are functionally independent with
respect to the $C^{(m)}_z$ (\ref{gn}) and, in general,  
$\{C^{(m)}_z,I_j\}\ne 0$. Hence we conclude that the Hamiltonian 
(\ref{si}) is superintegrable.

 Finally, we remark that a similar procedure can be carried out
for the second solution (\ref{sg}), thus obtaining another
superintegrable  Hamiltonian.

%%%%%%%%%%%%%%%%%%%%%%%%%%%%%%%%%%%%%%%%%%%%%%%%%%%%%%%

\sect{Angular momentum chains}

The connection between the
 $sl(2,\R)$ oscillator chains without centrifugal terms
($\otra_i=0$) and `classical spin' systems can be also
 extracted from the underlying coalgebra structure. As it was
shown in \cite{BR}, if we substitute the canonical realizations
used until now in terms of angular momentum realizations of the
same abstract
 $sl(2,\R)$ Poisson coalgebra, the very same construction will
lead us to a `classical spin chain' of the XXX Gaudin type on
which the non-standard quantum deformation can be easily
implemented. 

In particular, let us consider the $S$  realization of the
$sl(2,\R)$ Poisson algebra (\ref{fb}) given by  
\be
g_3^{(1)}=S(\jj)=\s_3^1 \qquad
g_+^{(1)}=S(\jp)=\s_+^1 \qquad
g_-^{(1)}=S(\jm)=\s_-^1
\label{ha}
\ee
 where the classical angular momentum variables $\sigma_l^1$
fulfil
\be
\{\sigma_3^1,\sigma_+^1\}=2\sigma_+^1\qquad 
\{\sigma_3^1,\sigma_-^1\}=-2\sigma_-^1\qquad
\{\sigma_-^1,\sigma_+^1\}=4\sigma_3^1 
\label{pb}
\ee
and are constrained by a given constant value of  the Casimir
function of
$sl(2,\R)$ in the form $c_1=(\s_3^1)^2 - \s_-^1 \s_+^1$. 

As usual, $m$ different copies of
 (\ref{ha})  (that, in principle, could have different values
$c_i$ of the Casimir) are distinguished with the aid of a
superscript
 $\s_l^i$. Then, the  $m$-th order of the  coproduct 
(\ref{primi}) provides the following realization of the
non-deformed $sl(2,\R)$ Poisson coalgebra :
\be
 g_l^{(m)}= (S\otimes \dots^{m)}\otimes
S)(\Delta^{(m)}(\s_l))=\sum_{i=1}^{m}{\s_l^i} 
\qquad l=+,-,3.
\label{hb}
\ee
Now, we  apply the usual construction and take ${\cal H}$ from
 (\ref{hsl}). As a consequence, the uncoupled oscilator chain
(\ref{fj}) with all $\otra_i=0$ is equivalent to the Hamiltonian
\be 
H^{(N)} 
=  g_+^{(N)}+\para g_-^{(N)}=\sum_{i=1}^{m}(\s_+^i + \para \s_-^i) 
\label{hc}
\ee
and the Casimirs $C^{(m)}$ read $(m=2,\dots, N)$: 
\be 
 C^{(m)}= (g_3^{(m)})^2 - g_-^{(m)} g_+^{(m)} 
  = \sum_{i=1}^{m}{c_i} + 
 \sum_{i<j}^{m} (\s_3^i \s_3^j- \s_-^i \s_+^j -  \s_-^j \s_+^i ).
\label{hd}
\ee 

Note that (\ref{hd}) are (up to constants) the classical
angular momentum analogues of XXX Gaudin Hamiltonians of the
hyperbolic type \cite{Gau}--\cite{EEKTb}. In other words, if we
consider the  $sl(2,\R)$ Casimir function as the dynamical 
Hamiltonian
${\cal H}=\jj^2 - \jm \jp$, such a Gaudin system can be
obtained through the coalgebra symmetry \cite{BR,BCR}. 
As a consequence, a non-standard deformation of the Gaudin system
can be now constructed through the deformed Casimir by taking
\be
{\cal H}=\jj^2  - \frac {\sinh z\jm}{z} \jp .
\label{hyu}
\ee
The deformed angular momentum realization
corresponding to $U_z(sl(2,\R))$ is: 
\bea
&&{\tilde g}_-^{(1)}=S_z(\jm)=\s_-^1 \qquad
{\tilde g}_+^{(1)}=S_z(\jp)=\frac {\sinh z \s_-^1}{z \s_-^1}
\s_+^1
\cr
&& {\tilde g}_3^{(1)}=S_z(\jj)=\frac {\sinh z \s_-^1}{z
\s_-^1}\s_3^1 
\label{ia}
\eea
where the classical coordinates $\s_l^1$ are defined on the cone  
 $c_1=(\s_3^1)^2 -\s_-^1 \s_+^1=0$, that is, we  are considering
the zero realization. It is easy to check that the $m$-th order
of  the coproduct (\ref{ga}) realized in the above representation
(\ref{ia}) leads to the following functions
\bea  
 && {\tilde g}_-^{(m)}= \sum_{i=1}^m \s_-^i\qquad
 {\tilde g}_+^{(m)}=\sum_{i=1}^m
 \frac {\sinh z \s_-^i}{z \s_-^i} \s_+^i  e^{z \kk_i^{(m)}(\s_-)
}\cr &&{\tilde g}_3^{(m)}=\sum_{i=1}^m
\frac {\sinh z \s_-^i}{z \s_-^i} \s_3^i  e^{z \kk_i^{(m)}(\s_-) }
\label{ib}
\eea
 that define the non-standard deformation of  (\ref{hb}).
Therefore, the $N$-th coproduct of the deformed Casimir gives
rise to the non-standard Gaudin system
\bea
  H^{(N)}_z&\equiv&C^{(N)}_z = ({\tilde g}_3^{(N)})^2  - \frac
{\sinh z{\tilde g}_-^{(N)}}{z} \, {\tilde g}_+^{(N)} \cr
&=&
\sum_{i<j}^N 
\frac {\sinh z \s_-^i}{z \s_-^i}\frac {\sinh z
 \s_-^j}{z \s_-^j} \, e^{  z \kk_{ij}^{(N)}(\s_-) } (\s_3^i
\s_3^j -
\s_-^i \s_+^j  - \s_+^i \s_-^j)
\label{id}
\eea
that, by construction, commutes with all lower dimensional Gaudin
Hamiltonians $C^{(m)}_z$ with $m<N$ and also with any of the
$N$-sites representation (\ref{ib}) of the generators of the
deformed algebra. This Hamiltonian is the angular momentum
counterpart to (\ref{gn}) for
$\otra_i=0$. 

If we recall the result coming from the standard deformation of
$sl(2,\R)$
\cite{BR}
\be
C_z^{(N)}= 
2\,\sum_{i<j}^{N}{\frac{\sinh
(\tfrac{z}{2}
\j_3^i)}{\j_3^i\,z/2}\,\frac{\sinh (\tfrac{z}{2}
\j_3^j)}{\j_3^j\,z/2}\,e^{z\,\kk_{ij}^{(N)}(\j_3)/2}\,\left(
\j_3^i\,\j_3^j - \j_1^i\,\j_1^j  -\j_2^i\,\j_2^j\right) }
\label{sss}
\ee
where $(\j_3^i)^2 - (\j_1^i)^2  -(\j_2^i)^2 =0$, we observe a
strong formal similarity with respect to (\ref{id}) since the
deformation can be interpreted in both cases as the introduction
of a variable range interaction in the model (compare (\ref{id})
and (\ref{sss}) with (\ref{hd})). However, within the
non-standard deformation the variable range factor is constructed
in terms of functions of $\s_-$ (note that
 $\jm$ is the primitive generator in this deformation), whereas
the standard one contains functions of
 $\j_3$ ($\jj$ is now the primitive one), and the geometrical
meaning of both coordinates is completely different.

%%%%%%%%%%%%%%%%%%%%%%%%%%%%%%%%%%%%%%%%%%%%%%%%%%%%%%%

\sect{Concluding remarks}

The systems here presented can be seen as basic examples of the
 implementation of integrable nonlinear interactions through
quantum algebras. We would like to recall again the universality
of this construction, that ensures the obtention of deformed
integrable systems by using non-trivial representations of any
quantum algebra with Casimir element.

It is interesting to stress that coalgebra symmetries are also
relevant at the undeformed level, since they account for the
integrability properties of known systems like the isotropic
$N$-dimensional oscillator and the Gaudin magnet. Note that these
 two systems are superintegrable and we have seen that, in the
first case, some choices of the deformed Hamiltonian do preserve
superintegrability. The question concerning the precise
characterization of the deformations and dynamical Hamiltonians
that follow this rule is an open question.

 Finally, although further analysis of the deformed dynamics of
these models is needed, two main features can be already be
pointed out: The  long-range nature of the interactions and their
 dependence on momenta.  The latter fact can be explored through
the generalized nonlinear  oscillators (\ref{ge}) that
 appear as the one-particle Hamiltonians $H^{(1)}_z$ and from
which the higher dimensional systems are constructed. At this
respect is interesting to recall that some relations between
generalized nonlinear oscillators and dissipation can be
established
\cite{UPVB,Chou}. In the same direction, some connections between
quantum algebras and this kind of phenomena have been already
envisaged \cite{IV}. Finally, notice that long-range interactions in
discrete systems can be linked to dispersive effects in the
continuum limit (see \cite{CGV} and references therein).

%%%%%%%%%%%%%%%%%%%%%%%%%%%%%%%%%%%%%%%%%%%%%%%%%%%%%%%

%\newpage

\bigskip
\bigskip

\noindent
{\Large{{\bf Acknowledgments}}}

\bigskip

\noindent The authors  have been partially supported by DGICYT
 (Project  PB94--1115) from the Ministerio de Educaci\'on y
Ciencia de Espa\~na and by Junta de Castilla y Le\'on (Project
CO2/399) and acknowledge discussions with Profs. O. Ragnisco, S.
Chumakov and G. Pogosyan.

%%%%%%%%%%%%%%%%%%%%%%%%%%%%%%%%%%%%%%%%%%%%%%%%%%%%%%%

\end{document}